\documentclass[a4paper,12pt]{article}%
\usepackage{amsmath}%
\usepackage{txfonts}%
\usepackage{amsfonts}%
\usepackage{amssymb}%
\usepackage{graphicx}
\usepackage{wrapfig}
\usepackage{color}
\usepackage{setspace}
\usepackage{esint}
\usepackage[rightcaption]{sidecap}
\usepackage{wrapfig}
\usepackage{lipsum}
\usepackage{caption}
\usepackage{longtable}
\usepackage[T2A,T1]{fontenc}
\usepackage[utf8]{inputenc}
\usepackage[russian,english]{babel}

\usepackage{etoolbox}
\makeatletter
\patchcmd{\@maketitle}{\begin{center}}{\begin{flushleft}}{}{}
\patchcmd{\@maketitle}{\begin{tabular}[t]{c}}{\begin{tabular}[t]{@{}l}}{}{}
\patchcmd{\@maketitle}{\end{center}}{\end{flushleft}}{}{}
\makeatother

\numberwithin{equation}{section}

\newcommand{\be}{\begin{equation}}
\newcommand{\ee}{\end{equation}}

\usepackage{anysize}
\marginsize{2cm}{2cm}{2cm}{2cm}
\usepackage{mathptmx}
-1in\relax 
\begin{document}

\begin{center}
\LARGE
\textbf{History of Prime Movers and Future Implications}
\normalsize
\vskip1cm
\begin{tabular}{l}
\hskip1cm
\Large Mikhail V. Shubov         \\
\normalsize \\
%University of New Hampshire       &
\hskip1cm University of MA Lowell    \\
%33 Academic Way                   &
\hskip1cm One University Ave,        \\
%Durham, NH 03824                  &
\hskip1cm Lowell, MA 01854           \\
%E-mail: marianna.shubov@gmail.com &
\hskip1cm E-mail: mvs5763@yahoo.com  \\
\end{tabular}
\end{center}

\begin{center}
  \textbf{Abstract}
\end{center}
\begin{quote}
Motive and electrical energy has played a crucial role in human civilization.  Since Ancient times, motive energy played a primary role in agricultural and industrial production as well as transportation.  At that time, motive energy was provided by work of humans and draft animals.  Later, work of water and wind power was harnessed.  During the 19$^{\text{th}}$ century, steam power became the main source of motive energy in USA and Britain. Modern transportation and industry depend on the work of heat engines that use fossil fuel.  A brief history of different sources of energy is presented in this work.  The energy consumptions in pre-industrial and industrial societies are calculated.  The lost opportunities for the Second Industrial Revolution (such as fast breeder reactors and thermonuclear power stations) are discussed.  The case that the Solar Power will become the main source of energy by the second half of this century is presented.
It is calculated that the Solar Power has the potential to bring about the new Industrial Revolution.  Based on material and energy resources available in the Solar System, it is demonstrated that the Solar System Civilization supporting a population of 10 Quadrillion with a high standard of living is possible.\\ \\
\textbf{Keywords:} motive energy, prime movers, Industrial Revolution, solar power, Solar System Colonization
\end{quote}

\section{Introduction}
Motive energy and mechanical work done by humans, animals, and machines has been one of the defining factors for Human Civilization.  At first, humans had to perform work without any assistance.  Since Early History of Humankind, work animals were used to carry loads, pull carts, and perform agricultural work. Since 3$^{\text{rd}}$ century BCE, water wheel power came into use \cite[p.9]{XRG1}.

During the XIXth century, there has been a tremendous growth in motive energy production.  At that point, steam engines were the main source of power \cite[p. 503]{HStat2}.  The growth of motive energy production enabled an unprecedented growth of industry and income per capita.  Rapid growth of motive energy production continued up to about 1970.

Currently, motive energy production is stalled.  Modern civilization relies on fossil fuels to produce motive energy.  This energy production can hardly expand.  Many scientists believe that Solar Power will become the main source of energy within a few decades \cite{SRev01,SRev02,SRev03}.  In this work we present a case, that Solar Power can not only replace fossil fuels as the main source of energy, but also enable growth of energy production by a factor of 50 to 150.  This energy growth would bring the Second Industrial Revolution and great increase in gross domestic product (GDP) per capita.  The final stage of Human Civilization would be colonization of Solar System.  That would expand energy production by a factor of about 100 billion.  As we discuss in Section 8, Solar System Civilization would be able to support a population of 10 quadrillion people.

At this point, we present a strict definition of \textbf{prime movers} and \textbf{motive energy}.  A prime mover is any engine producing mechanical power for a vehicle, manufacture, or electricity generation.  Work animals are counted among prime movers, which is relevant for past centuries.  In 1850, about 60\% of motive energy in USA was produced by work animals \cite[p. 503]{HStat2}.  Motive energy is the total energy produced by prime movers.  It also includes all electric energy from any source.

Many sources dealing with modern energy production and consumption confuse motive energy with heat energy.  Thus, electric energy produced by nuclear, hydroelectric, wind, or solar power is counted at the same rate as potential chemical energy in petroleum or natural gas.  This is absolutely wrong, since most modern engines convert the energy present in fuel into motive or electric energy at 37\% efficiency \cite[p.213]{EEff}.

\section{Pre-Industrial Age}
Working animals have been the most important source of motive energy in pre-Industrial world.  Animals were used for plowing, transportation, and driving mills.  Water and wind power were the other major sources of motive energy.

Water wheels originated in Syria in 3$^{\text{rd}}$ century BCE \cite{WWheel}.
Water wheel powered hammers became common in Italy in the 1$^{\text{st}}$ century CE \cite{THammer1}.  They were common in China at the same time \cite[p.183]{THammer2}.  Water powered saw mills became common by 11$^{\text{th}}$ century \cite{Smill}.  Fulling mills appeared in 11$^{\text{th}}$ century \cite[p.14]{MMMach}.  During the Middle Ages, water wheels began powering bellows for blast furnaces, tool sharpening wheels, drills for making cannons, chopping mills for making paper, and lathes  \cite{WaterWheel}.

In order to estimate the energy production in pre-Industrial World, we must have an estimate for the time worked by each prime mover.
In a developed pre-Industrial society there was about one water wheel per 300 inhabitants.  This number definitely varied by society.  Each wheel developed an average of 3.7 $kW$ and worked 2,200 hours per year \cite[p.7961]{PMO2}.  Thus water power provided an average of 27 $kWh$ per year per inhabitant.
By far the greatest contribution of wind power was for sailing vessels.
An average ship sailed 3,500 hours per year.  Average work performed by wind on sailing vessels in developed pre-Industrial societies is approximately equal to 33 $kWh$ per year per inhabitant \cite[p.7959]{PMO2}.  Once again, there was variance among societies.

Draft animals provided most work in pre-Industrial world.
In a developed pre-Industrial society such as USA in mid nineteenth century, there was about 0.25 $hp$ of working animal power per capita \cite{HStat2}.  This could be a horse or two bulls per four inhabitants.  Obviously, number of animals per capita also varied from place to place.
Animals can work up to 6 hours per day.  From data in \cite[p.11]{HpH01}, it follows that an average draft animal in USA 1850 did an equivalent of 900 hours per year full intensity work.  In modern India, an average draft animal works 600 hours per year \cite{DAN01,DAN02}.  According to other sources, 660 hours per year is the normal workload for an animal, while 1,200 hours per year can only be sustained by a camel \cite{Animals1}.
Based on the data above, draft animals provided an average work of 110 $kWh$ to 160 $kWh$ per year per inhabitant.

A good estimate for the total amount of motive energy per inhabitant in pre-Industrial society is 200 $kWh$ per year, of which 140 $kWh$ came from work animals, 30 $kWh$ from water power and 30 $kWh$ from wind power.  In 2017, US motive energy consumption was 20,400 $kWh$ per capita. Of that energy, 12,600 $kWh$ per capita is electricity \cite[p.69]{WEnrg18} and about 7,800 $kWh$ per capita is gas engine work \cite[p.144]{Trnsp}.

The combination of plant photosynthesis and animal metabolism can be considered the Nature's way of converting solar power to motive energy.  This way is very inefficient.  Most crops convert only about 0.3\% of the energy of sunlight into food calories \cite{korm}.
Work animals are not efficient engines.  From the data presented in \cite[p.7958]{PMO2} and \cite[p.11]{HpH01}, it follows that only 6.5\% of energy in draft animal feed was converted to useful work.
Other sources studying modern India estimate draft animal efficiency at 4.0\% \cite{korm1} to 5.0\%
\cite{korm2}.
Overall, 0.02\% of solar energy is converted into motive work.  A modern 16\% efficient photovoltaic cell has 800 times greater efficiency.

Some modern researchers suggested growing energy rich crops and using them to produce diesel fuel.  The best choice among current crops is palm oil.  The system would be 0.45\% efficient \cite[p.23]{algae}.  A system using algae can be 1.5\% efficient \cite[p.6]{algae1,algae2}.  This still requires very much work, and is still vastly inferior to photovoltaic cells.  Algae is best suited for producing feed for pigs, poultry, cattle and fish \cite[p.23]{algae1}, but it can not compete with photovoltaic cells in harvesting solar power.

\section{Steam Power}
In 1698, Thomas Savery invented a steam pump.   Savery pumps did not work autonomously -- they required an attendant switching two valves at regular intervals. Savery pumps had a power of 0.7-0-8 $kW$.  These pumps were mostly used to pump water out of mines \cite[p.4]{duty2}.
The first practical steam engine was invented by Thomas Newcomen is 1712 \cite{Newcomen}.
First Newcomen engines had a power of 4 $kW$, while some later ones produced up to 56 $kW$ \cite[p.7]{duty2}.

Steam engine was further improved by James Watt in 1760s and 1770s \cite{Watt}.
By 1800, 496 Watt engines have been produced.  These engines had 5-10 $kW$ power \cite[p.9]{duty2}.
In 1797, Robert Trevithick invented the first  \textbf{high pressure steam engine} \cite[p.6]{duty2}.
In a high pressure stem engine, steam expands in a cylinder and thus performs work.  All previous engines were \textbf{atmospheric steam engines}.  In an atmospheric steam engine, steam condenses and creates partial vacuum within a cylinder.  The atmosphere does work on a piston by pushing it inside \cite{Newcomen}.
In 1849, the steam engine was further improved by George Corliss \cite{Corliss}.  In 1862, Porter and Allen developed a high speed stem engine \cite{PorterAllen}.

Efficiency of steam engines improved over time.
Savery pumps had efficiency below 0.5\% \cite[p.4]{duty2}.
Original Newcomen engines had efficiency of 0.5\%.
Later Newcomen engines improved by Smeaton had efficiency of 1\% \cite[p.7]{duty2}.
Watt engines had efficiency of 2\% to 3\% \cite[p.87]{duty1}. By 1840s, top steam engines had efficiency of 12\% \cite[p.15]{duty2}.
When steam engines were used to drive factory machinery, most energy was lost in  transitions.  The efficiency of factories, which included both the engine and the mechanical transitions was much lower.  In in USA 1900, factory efficiencies averaged 4\% \cite[p. 354]{HStat2}.

The number of steam engines and their cumulative power grew rapidly.  By 1800, there were about 600 steam engines in the World, mostly in Britain.  By 1810, the number of steam engines grew to about 5,000 \cite[p.16]{duty2}.  By 1810s, steam was still not a significant source of motive energy -- like solar power is still not a significant source of electricity in 2020.
By 1840,    570 $MW$ steam power was installed in USA and    650 $MW$ in Europe.
By 1870,  4,200 $MW$ steam power was installed in USA and  8,800 $MW$ in Europe.
By 1896, 13,500 $MW$ steam power was installed in USA and 30,200 $MW$ in Europe \cite[p.16]{duty2}.
Hopefully, solar power will be the primary source of energy at the end of this century.  For now, figures for the years 2040, 2070, and 2096 are not available.

Steam power played a key role in Great Britain's Industrial Revolution.  By 1850, steam power was used in a wide variety of manufacturing.   It was used in food industry, tobacco manufacture, textile industry, lumber and wood products, paper production, chemical industry, and metal working \cite[p. 458]{PWR}.

Steam turbines had been introduced in 1884 by Sir Charles Parsons \cite{CPars}.  First steam turbines were very inefficient and had low power.  By 1900, a 1.3 $MW$ turbine was built.  By 1907, a 13 $MW$ turbine was built. The first gigawatt turbine was built in 1965.  Steam turbine efficiency grew from 12\% in 1900 to 30\% in 1930 to 42\% in 1973 \cite[p. 38-39]{duty2}.  Most electrical energy in 2019 is generated by power plants using steam turbines.

Steam turbines are likely to have an important role to play during Space Age and colonization of the Solar System.  In outer space, energy can be generated by turbines using potassium vapor as the working fluid. The cycle is closed.  Potassium is heated either by nuclear or concentrated solar energy \cite{KTurbo1,KTurbo2,KTurbo3}.

\section{Fossil Fuel Era}
During the First Industrial Revolution, rapid growth of energy production was enabled by the use of the heat engines powered by fossil fuel.  These heat engines could produce much more power than waterwheels, windmills, and work animals.

Between 1849 and 1923, the total power of engines installed in industry grew 68 times \cite[p. 30]{PMO1}.  Between 1849 and 1955, the total power of prime movers used in American Industry and transportation grew by a factor of 840.  This factor overestimates the actual growth of motive energy production.  In 1955, 93\% of all power of prime movers was in automobile engines \cite[p. 503]{HStat2}.  On average automobiles work only a small fraction of time, and do not use their full power.

According to detailed studies, 6.7 billion $kWh$ of motive energy has been produced in USA 1850 \cite[p.11]{HpH01}.  The total motive energy produced in 1956 can be calculated from mineral fuel production. The heating value of mineral fuel produced in USA 1955 is 11.0 trillion $kWh$ \cite[p. 354]{HStat2}.  About 42\% of this value has been converted to motive energy \cite[p.70]{HpH01} at an average efficiency of 28\% \cite[p. 507]{HStat2}.  Overall, 1.3 trillion $kWh$ of motive power has been produced in USA, 1955.  Between 1850 and 1956, the total energy motive energy produced in USA increased by a factor of 195.  Between 1890 and 1980, GNP and energy consumption in USA have been closely correlated \cite[p.6]{HStat3}.  Among modern Nations, energy consumption per capita is almost proportional to GDP per capita to the power 0.78 \cite[p.20]{XRG1}.

In USA 1900 to 1955, total mineral fuel production grew by a factor of 5.0.  During the same time, the average efficiency of electric power production grew from 4\% to 28\% \cite[p. 354]{HStat2}.  By 2011, the average efficiency has grown to 35\% \cite[p.326]{NUCL01} -- which indicates very slow progress.  The efficiency of an electric power production is the product of the prime mover efficiency, the electric generator efficiency, and grid transmission efficiency.  Generator efficiencies are generally above 90\%.  By 1911, alternating-current generators had efficiencies of 94\% to 96\% \cite[p.43]{GEff01}.  Efficiency of electric power use in industry has also undergone significant improvement over the last century \cite{ElEff}.

Between 1973 and 2017, global fossil fuel consumption grew from 71 trillion $kWh$ to 162 trillion $kWh$ in thermal energy equivalent \cite[p.8]{WEnrg18}.  During the same time, electric power generation efficiency grew from 32.7\% to 37.0\% \cite[p.213]{EEff}.  Combining the aforementioned data, we conclude that motive energy equivalent grew from 23 trillion $kWh$ to 60 trillion $kWh$ between 1973 and 2017.  Electric power generation itself grew from 6.1 trillion $kWh$ to 25.6 trillion $kWh$ during these years \cite[p.30]{WEnrg18}.

Sustaining economic growth based on fossil fuels is impossible.  Increasing the consumption of fossil fuels will lead to their depletion.  Increasing the efficiency of prime movers is a slow and expensive process.

\section{Motive Energy in Transportation}
Work animals have been used for pulling carts for about four millennia \cite{Neolit}.  Wind power has been used to propel sailing vessels since Ancient Egyptian times \cite{AEgypt}.

The real proliferation of steam transportation came only with the introduction of railroads and steam locomotives.  Richard Trevithick, who invented the high pressure steam engine built the first railroad locomotive in 1804.  It pulled five wagons weighing 10 $tons$ for a distance of 16 $km$ at a speed of 8 $km/h$ \cite[p.10]{duty2}.

First railroads in Britain and USA were built in late 1820s.  By 1840, USA contained 2,800 miles of railroads, by 1850 -- 9,000 miles, by 1860 -- 30,000 miles and by 1900 almost 200,000 miles.  Railroad development was sped up by a fast growth in steel production during the second half of XIXth Century \cite[p. 133]{USOld}.
Speeds which have been unimaginable earlier became reality.  By mid 19$^{\text{th}}$ century, train speeds of up to 100 $km/h$ became common \cite[p.19]{duty2}.

Number of passenger-miles rose more rapidly than the length of railroads.  It rose from 470 million passenger-miles in 1849 to 1.9 billion passenger-miles in 1859 and 12 billion passenger-miles in 1890 \cite[p. 585]{RRail}.  The first diesel locomotive appeared in 1925.  By 1957, diesel locomotives were 10 times as numerous as steam locomotives \cite[p.429]{HStat2}.  Even though passenger cars have displaced trains as the primary mode of passenger transportation since 1920s, trains remain important in freight transport.  The amount of freight moved by train tripled between 1960 and 2006 \cite[p.9]{Trnsp}.

In USA, first steam ship went afloat in 1809.  By 1840, 10\% of all American ships were steam-powered.  In 1893, for the first time, steam ships outnumbered sailing ships \cite[p.445]{HStat2}.

Electric Streetcar Revolution started in 1888 and spread rapidly \cite{Spargue}.  By 1902, there were almost 60,000 electric street cars in USA, which carried 4.5 Billion passengers that year \cite[p.6]{StreetCar}.  The street cars travelled 1.1 Billion miles \cite[p.12]{StreetCar}.

Automobiles were first proposed by Leonardo da Vinci \cite[p.7]{SelfDriving}.  In 1769, Nicolas-Joseph Cugnot built the first steam-powered car \cite[p.8]{SelfDriving}.  During 1830s, Walter Hancock built three steam-powered passenger buses which were much more successful and less expensive than contemporary horse-drawn buses \cite{SteamRoad}.  The buses travelled at an average speed of 10 mph.  They travelled an average of 53 miles per day.  Each bus carried an average of 30,000 passengers and performed 180,000 passenger-miles per year \cite[p. 77]{SBus}.  Walter Hancock planned to expand his omnibus line to about 80 steam carriages \cite[p. 86]{SBus}.  In 1831, H.T. Alken predicted that steam automobiles would soon displace horse transportation \cite{SCoach}.

Both Walter Hancock's plan and  H.T. Alken's prediction failed.  For many decades, the automotive age did not come.  Some technological projects are impossible at the time of their conception.  Nevertheless, almost all of these projects become possible as technology advances.  Automotive age did come.  In 1960s and 1970s, many futurists believed that Space Age is coming soon \cite{SAge1,SAge2,SAge3}.  It still has not come.  Success in once abandoned projects should give us hope.

The first gasoline-powered car was first built in 1885 \cite[p.8]{SelfDriving}.  At first car production was slow.  Henry Ford build an assembly line which produced Model T cars in large numbers \cite{FModT}. In USA, the number of automobiles rose from 8,000 in 1900 to 458,000 in 1910, 8.1 million in 1920, 23 million in 1930, and 56 million in 1957  \cite[p.462]{HStat2}.  In 2012, there were 254 million motor vehicles in USA \cite[p.9]{Trnsp}.

The next great challenge in transportation technology is the ability to transport astronauts and payload into outer space at reasonable cost.  The first successful space launch took place on October 4, 1957 -- a Soviet satellite named Sputnik was placed in orbit \cite{Sputnik}.  In 1961, the first astronaut named Yuri Gagarin went to space \cite{Gagarin}.  American Lunar Expedition took place in 1968.

Unfortunately, \textbf{launch costs}, which are the costs of placing payload into Earth's orbit remained high.
Up to 2010s launch costs remained at an average of \$18,500 per $kg$ up to about 2010 \cite[p.8]{LCost}.  A breakthrough in launch cost reduction was accomplishes by SpaceX company.  By 2009, their Falcon 9 rocket delivered payload to LEO for \$2,700 per $kg$.  The next step was the introduction of the reusable first stage.  On December 21, 2015, Space X made a huge step in History when the first stage of Falcon 9 spacecraft returned to the launching pad \cite[p.1]{SLVD}.  During 2016, SpaceX has successfully landed six first stage boosters \cite{Falcon2}.  By July 2019, there have been 34 successful first stage returns out of 40 attempts \cite{Falcon3}.  By 2018, SpaceX was offering LEO delivery at \$1,400 per $kg$ via Falcon Heavy \cite[p.8]{LCost}.

Many engineers promised drastic reduction of launch costs for decades.  At this point we can not predict the future development of technology and launch cost reduction.  It is possible that True Space Age and colonization of Solar System will occur during the next Energy Revolution.

\section{Nuclear Power -- a Lost Chance}

The first nuclear power plant in USA was built by 1957.  By 1970, 20 nuclear power plants operated.  By 1980 there were 71 nuclear power plants, and 112 nuclear power plants by 1990 \cite[p.271]{NUCL01}.  Electricity generation by the nuclear power plants increased even more rapidly.  In 1957, the nuclear power plant generated 0.2 billion kWh.  In 1970, nuclear power plants generated 22 billion kWh.  These plants generated 250 billion kWh in 1980 and 577 billion kWh in 1990 \cite[p.273]{NUCL01}.  Continued growth of nuclear power production could have started the new Industrial Revolution.  Nuclear Power Revolution could have started in 1990s and continued during the first decades of this Century.  Unfortunately, the Nuclear Power Revolution came to an abrupt end before it really started.  Nuclear share of total net generation has not changed much since 1988 \cite[p.273]{NUCL01}.

In order to understand the fizzling of Nuclear Power, we must have basic understanding of nuclear reactors.  There are several types of nuclear reactors. The author's paper \cite{MyMasters} was on the subject of Accelerator Breeder Reactors.  Other reactor types relevant to this article are Thermal Reactors and Fast Breeder Reactors discussed in paragraphs below.

In all nuclear reactors, a chain reaction of nuclear fission is sustained.  When a \textbf{fissile nucleus} absorbs a neutron, it is likely to undergo a nuclear fission event.  Examples of fissile nuclei are $^{233}$U, $^{235}$U, and $^{239}$Pu.  A nuclear fission produces several secondary neutrons.  The average number of secondary neutrons produced depends on the energy of absorbed neutron and the nucleus undergoing fission.  Generally, the average number of secondary neutrons per fission is 2.4 to 2.9.  Some of the secondary neutrons are lost, while others cause further fission reactions.  In a sustained nuclear fission, the number of neutrons absorbed is about the same as the number of neutrons produced.  The total neutron flux changes very little over time.

In Thermal Nuclear Reactors, the neutrons are slowed down before they cause a nuclear fission.  Neutrons can be slowed down by multiple collisions with nuclei.  Thermal reactors are by far the most common ones.  In Fast Breeder Reactors, the chain reaction is sustained by fast neutrons.  In Accelerator Breeder Reactors, the nuclear chain reaction is not self-sustaining.  This reaction is sustained by an external source of neutrons.  That source of neutrons consists of a uranium target subject to a stream of super energetic protons.  These protons have energy of about 1 GeV.  This proton stream is produced by an accelerator.  Whenever a super energetic proton strikes a heavy nucleus it causes the nucleus to disintegrate into many light fragments and neutrons \cite[p.8-13]{MyMasters}.

All reactors consume fissile nuclei such as $^{235}$U, $^{233}$U, and $^{239}$Pu.  Most reactors  also produce fissile nuclei from \textbf{fertile nuclei}.  Examples of fertile nuclei are $^{232}$Th and $^{238}$U.
When $^{232}$Th absorbs a neutron, it becomes $^{233}$Th, which decays to $^{233}$U -- a fissile nucleus.
When $^{238}$U absorbs a neutron, it becomes $^{239}$U, which decays to $^{239}$Pu -- a fissile nucleus.

In Thermal Nuclear Reactors, consumption of fissile nuclei greatly exceeds production of fissile nuclei from fertile nuclei.
In Fast Breeder Reactors, and more so in Accelerator Breeder Reactors, production of fissile nuclei from fertile nuclei considerably exceeds consumption of fissile nuclei.
As a result, Thermal Nuclear Reactors must use the resources of fissile nuclei.  Fast Breeder Reactors and more so in Accelerator Breeder Reactors can use the resources of fertile nuclei.  Fertile nuclei are much more common in nature than fissile nuclei.  The only naturally occurring fissile isotope is $^{235}$U, which makes up 0.7\% of all uranium found in nature.  The rest of natural uranium is fertile $^{238}$U \cite[p.6]{MyMasters}.  In terms of global energy reserves, $^{235}$U contains 21 times less energy than coal \cite[p.17]{WorldEnergy}.  Reserves of fertile isotopes are virtually unlimited.  A ton of average rock contains 18 $g$ of thorium, and 3 $g$ of uranium. That is an energy equivalent to 45 $tons$ of coal \cite[p.8]{MyMasters}!

Thermal Nuclear Reactors may be useful for limited applications.  They are useful for marine propulsion \cite{korabli}.  Nevertheless, they can not replace fossil fuel as the main source of energy due to lack of sufficient resources of $^{235}$U.  In the author's work on nuclear reactors \cite{GCR}, a case was made that thermal nuclear reactors could be very useful for space propulsion .

The author also made a case against proliferation of thermal nuclear reactors on Earth -- $^{235}$U consumed in these reactors will deplete a fuel resource needed for space transportation \cite[p. 102]{GCR}.    Total resources of uranium producible at \$130 per $kg$ or less is 6,140,000 $tons$ \cite[p.15]{U2018}.  In 2019, Thermal Nuclear Reactors consumed $^{235}$U contained in 67,000 $tons$ of uranium \cite{WorldNuclear}.  By the time $^{235}$U will be needed for space exploration, most uranium resources may be depleted.

No Accelerator Breeder Reactors have been built.  In December 2019, there are 444 nuclear reactors in the World with total power of 395 $GW$ \cite{WorldNuclear}.  There are also 6 Fast Breeder Reactors in the World with total power of 2 $GW$ \cite{WorldNuclear1}.  Fast Breeder Reactors held a promise of providing unlimited energy supply \cite{LMBBR1}.
%As I was working on my Masters in 1999-2000, I was certain that by 2020, Nuclear Energy Revolution powered by Accelerator Breeder Reactors and Fast Breeder Reactors would be underway.

Nuclear Fusion power also seemed very promising.  According to a 1960 report, there should be about 250 nuclear fusion power plants in Europe in 20 years \cite{NFusion}.  Some people are still optimistic about this source of energy, while others have given up hope.  One of the main reasons why Nuclear Fusion did not succeed is that it has received very little funding.
Between the years 1975 and 1982, the average annual budget for fusion power in USA was \$1 billion per year, after which the funding fell rapidly \cite{NFBudget1}.  Between the years 2000 and 2012, the average annual budget for fusion power in USA was \$300 million to \$400 million per year \cite{NFBudget2}.  According to a 1976 plan for development of nuclear fusion power, these levels of funding would never achieve result \cite[p.12]{FPlan}.  In Europe, a giant thermonuclear power station called ITER is being constructed.  It's total cost of \$22 Billion is covered by 35 Nations.  It is supposed to start working in 2035 \cite{NFBudget3}.

In the author's opinion, Nuclear Fusion based Energy Revolution would have succeeded if it had more funding.  Many experts agree \cite{NFFund1,NFFund2,NFFund3,NFFund4}.  Had funding for Fusion Power been at least \$30 Billion per year since 1980, it is likely that Fusion Power Revolution would have started by the turn of the century.

\section{Future Prospect -- Solar Power Revolution}
Energy production has little chance for growth in the coming decades.  Almost all of the energy comes from fossil fuels, which are in a very limited supply.  Total reserves of fossil fuel can sustain 82 years of use at current rate \cite{WEnrg18,WorldEnergy}.  The world may contain up to 17 trillion tons of hard coal \cite[p.28]{WorldEnergy}, but using this reserve is likely to cause enormous global warming.

The technology which has a potential for totally transforming energy production is harvesting of Solar Power.  In order to understand the possible impact of Solar Power Revolution, we must compare the amount of motive power produced in Modern World to the amount of motive power which can be produced by Solar Power.  As we have mentioned earlier, global energy consumption is equivalent to 60 trillion $kWh$ of motive energy per year.  If all of Earth's deserts are covered with 16\% efficient photovoltaic cells, then the total electricity production would be 5.0 quadrillion $kWh$ per year.

A very interesting technology is Floating Solar Power -- solar power stations floating on water.  Currently, only 0.4\% of all Photovoltaic power is produced by floating solar power stations \cite{FSolar}.  By the end of Solar Power Revolution, Floating Solar Power may become the main energy source.  If 20\% of World Ocean is covered by 16\% efficient photovoltaic cells, then the total electricity production would be 10.0 quadrillion $kWh$ per year.  This is twice as much as we can obtain from deserts.  All deserts and 20\% of ocean can bring 15 quadrillion $kWh$ per year, which is 250 times greater than modern motive power production.

In 2017, worldwide, Solar Power produced about 2.5\% of global electricity and 0.9\% of global motive energy.  That year 531 billion $kWh$ of electricity was produced by solar power \cite[p.76-77]{PV17}.  Solar Power production has been growing by an average of 44\% per year since 1992.  It has been growing by an average of 32\% between 2012 and 2017 \cite[p.82]{PV17}.

At the time, the cost of installed photovoltaic power fell rapidly.  Between 2010 and 2018, the cost of installed solar power for utility-scale stations fell from \$4.63 per $Watt$ to \$1.06 per $Watt$ \cite[p.viii]{PVCost}.  During the same time, the prices of solar modules themselves dropped from \$2.47 per $Watt$ to \$0.47 per $Watt$.  The main breakthroughs came between 2010 and 2013 and in 2016 \cite[p.43]{PVCost}.  By December 2019, most module prices fell to \$0.28 per $Watt$.  Electric energy produced by Solar Power Stations has an average production cost of 5 cents per $kWh$.  Cost decrease has surpassed the 2020 target \cite[p.39]{PVCost}.  Between 2010 and 2018 the average efficiency of the new photovoltaic modules installed in utilities in California grew from 13.8\% to 19.1\% \cite[p.5]{PVCost}.

If the growth rate of 20\% per year can be sustained for 20 years, then Solar Power would produce most of electric energy by 2040.  That year about 40 trillion $kWh$ electric energy should be produced by Solar Power.  If the energy production by other prime movers will remain relatively unchanged, the global motive energy production in 2040 should be about 100 trillion $kWh$.  The most likely scenario is that after that Solar Power production will continue to grow.  This will mean the growth of overall power production.  This will likely drive the Second Industrial Revolution.  The growth of Solar Power will continue until it will reach the natural limit of 15.0 quadrillion $kWh$ described above.

How long will Second Industrial Revolution take?  Obviously, we have no way of knowing.  Most past predictions about the present did not come true.  Nevertheless, we can make a judgement based on historical precedent.  As we have mentioned earlier, between 1850 and 1956, the total energy production by prime movers in USA grew by a factor of 210 \cite[p.507]{HStat2}.  This corresponds to a growth rate of 66\% per decade.  If global production of motive energy grows at the same rate during the Second Industrial Revolution, then it will take from 2040 to 2140 for motive energy production to grow from 100 trillion $kWh$ to 15.0 quadrillion $kWh$.  Obviously, we can neither rule out faster nor slower growth.  Hopefully, the Solar Power Revolution will not fizzle like Fast Breeder Reactors and Fusion Power.  Only time will tell.

\section{The Final Frontier}

Colonization of the Solar System is the Final Frontier for Humankind.  Resources contained within the Solar System are vastly greater then resources available within Earth's crust.  The total solar energy available in space exceeds the solar energy available on Earth by a factor of a billion.

The Solar System will provide a new home for most humans, even though Earth will remain an important cultural center.  Humans will live on billions of large habitats orbiting the Sun.  Each of these habitats will harvest solar energy.  Each habitat will produce all necessary food, drinking water, and oxygen needed for humans and animals.  Some goods will be produced on specialized factory habitats and distributed to other habitats.  This concept is called the Dyson Sphere \cite{Dyson}.  The concept of Solar System Civilization was first envisioned by Konstantin Tsiolkovsky in 1903 \cite{Tsialkovski}.  During 1970s, many elaborate models of Solar System Civilization were published \cite{habitats,1974}.

For the rest of this Section, we use the term \textbf{Exaton}, which is $10^{18}$ $tons$. The Asteroid Belt contains about 3 $Exatons$ of material composed of metal silicates, carbon compounds, water, and pure metals \cite{ABM}.  Most of asteroids are of a carbonaceous type \cite{Asteroids}.  Carbon is very useful for production of food for space travelers, fuel for propulsion within space, and plastics for space habitat structures.  High quality steel is also an abundant resource in space. For example, asteroid 16 Psyche contains $10^{16}\ tons$ of nickel-rich steel \cite{Psycho}.  Initially, asteroidal material would be sufficient for construction of space-based habitats.  Additional material for comfortable habitats can be obtained from Mercury, satellites of gas giant planets, and Kuiper Belt objects \cite{Kuiper}.  Kuiper Belt contains about 120 $Exatons$ of material -- mainly water, ammonia, and carbon compounds \cite{MKuiper}.  Planet Mercury contains 330 Exatons of material composed of metal silicates, carbon compounds, and pure metals \cite[p.14-2]{crc}.  Satellites of Jupiter and Saturn contain at least 10 Exatons of water and hydrocarbons \cite[p.14-4]{crc}.  Given the data above, it is possible to construct a total habitat space of 100 Exatons.  It has been estimated that Solar System resources can easily sustain a population a million times greater than the global population of today \cite{skymine}.  Each inhabitant will have space provided by 10,000 tons of structure.  The mass of modern luxury cruise liners can be approximated by multiplying 85 $tons$ by the number of cabins \cite{CruiseShips}.  Space habitats will have about 120 times more structural material per inhabitant, and habitat material will be more advanced.  This will provide material standard of living suitable for Solar System Civilization.

As we have discussed in this article, the most important resource for industry and civilization is energy \cite{kard, E1}.  Sunb's thermal power is $3.86 \cdot 10^{26}\ W$ \cite[p.14-2]{crc}.  A future civilization, which would harvest 1\% of that power with 15\% efficiency, will have energy production of $5 \cdot 10^{24}\ kWh/year$.  With Solar System Civilization being a home to about $10^{16}$ inhabitants, the motive energy consumption per capita would be 500 million $kWh$ per year.  As we have mentioned in Section 2, energy consumption in USA 2019 is 20,400 $kWh$ per year per capita -- almost 25,000 times less.  Nevertheless, life in a space habitat would require much more energy.  People at that time will likely view our material standards of living as rudimentary and poor.

When will Solar System Colonization take place?  In the author's opinion, technology to start colonization of Solar System existed since 1970s.  Many contemporary experts agreed \cite{habitats,1974}.  Elon Musk believes that colonisation of Solar System can start in 2020s \cite{Musk1,Musk2,Musk3}.  Each new invention and technology makes initial steps of Solar System Colonization more feasible.  The new Energy Revolution should create both capital and improve technology for Solar System Colonization.

We do not know when the Solar System will be colonised, but we can look for historical precedent.  Maritime technology of Ancient World may have been sufficient to sail to America \cite{RUSA}.  Leif Erikson discovered America in the beginning of the 10$^{\text{th}}$ century \cite{LErikson}.  Possibly, the Vikings could have started colonization of North America in 11$^{\text{th}}$ century.  Colonization of South America began after Columbus' discovery of the continent \cite{Columbus}.  If colonization of America did not start at that time, it definitely would have started in the 17$^{\text{th}}$ or 18$^{\text{th}}$ centuries.  As for colonization of the Solar System, only time will tell.

\end{document}